\documentclass[5p]{elsarticle}
\usepackage{graphicx, amsmath, amssymb, aas_macros}
\usepackage{natbib, xfrac}
\usepackage{hyperref}
\usepackage{bm, color}
\graphicspath{{}{./fig/}{./png/}}

\newcommand{\Fig}[1]{Figure~\ref{#1}}
\newcommand{\Eq}[1]{Equation~(\ref{#1})}
\newcommand{\EQ}{\begin{equation}}
\newcommand{\EN}{\end{equation}}
\newcommand{\vv}{\mbox{\boldmath $v$} {}}
\newcommand{\uu}{\mbox{\boldmath $u$} {}}
\newcommand{\Sec}[1]{Section~\ref{#1}}

\biboptions{authoryear}

\date{\today,~ $ $Revision: 1.1 $ $}
\begin{document}

\begin{frontmatter}

\title{Semarkona: Lessons for chondrule and chondrite formation}

\author[AIH]{Alexander Hubbard\corref{cor1}}
\ead{ahubbard@amnh.org}

\author[DES]{Denton S. Ebel}
\ead{debel@amnh.org}

\cortext[cor1]{Corresponding author}

\address[AIH]{Department of Astrophysics, American Museum of Natural History, New York, NY 10024-5192, USA;
Phone: 212-313-7911}
\address[DES]{Department of Earth and Planetary Sciences, American Museum of Natural History, New York, NY 10024-5192, USA;
Phone: 212-769-5381}

\begin{abstract}
We consider the evidence presented by the LL3.0 chondrite Semarkona, including its chondrule fraction, chondrule
size distribution and matrix thermal history.  We show that no more than a modest fraction of the ambient matrix material in the Solar
Nebula could have been melted into chondrules; and that much of the unprocessed matrix  material must have been filtered out
at  some stage of Semarkona's parent body formation process.  We conclude that agglomerations of many
chondrules must have formed in the Solar Nebula, 
which implies that chondrules and matrix grains had quite different collisional sticking parameters.  Further, we note that the absence
of large melted objects in Semarkona means that chondrules must have exited the melting zone rapidly, before the chondrule
agglomerations could form.  
The simplest explanation for this rapid exit is that chondrule melting occurred in surface layers of the disk.  The newly formed, compact,
chondrules then settled out of those layers on short
time scales.
\end{abstract}


\end{frontmatter}


\section{Introduction}

There have been many recent advances in the field of planet formation, including an improved understanding
of the earliest stage of growth where subcomponents are held together by electrostatic forces and chemical bonds.
This stage starts with sub-micron interstellar dust and ends with planetesimals held together
by gravity.  Numerical simulations and experiments have allowed us to probe this coagulation regime
\citep{2010A&A...513A..56G,2013ApJ...776...12P}, 
and have lead to the confirmation of collective behavior such as the Streaming Instability 
\citep[SI,][]{2005ApJ...620..459Y,2007Natur.448.1022J}.

We apply this recent understanding to the chondritic meteorite Semarkona.  An LL3.0 meteorite \citep{2005M&PS...40...87G},
 Semarkona experienced very little
parent body alteration, which means that it is an excellent record of the solids in the Solar Nebula.  Of particular note
is the fact that Semarkona is mostly made of chondrules about $0.5$~mm in diameter, with the remainder
being fine-grained matrix \citep{2014LPI....45.1423L}. While these chondrules are certainly
small from an every-day perspective, we will show that they are also too small to fit well with our current understanding of planetesimal formation.
This is even more troubling because, as an LL chondrite, Semarkona's chondrules 
are relatively large \citep{2006mess.book...19W}.
Constructing a theory for the formation of Semarkona's parent body is further complicated by
the low temperatures recorded in the matrix.

These difficulties mean that Semarkona's components put significant constraints on models of the Solar Nebula and the earliest
stages of planet formation, while those models conversely constrain our interpretations of laboratory investigations of Semarkona.
We take some early steps in combining the laboratory data with analytical and numerical studies of dust dynamics
in the Solar Nebula, considering
formation scenarios in which chondrules were made by melting free-floating clumps of dust, and subsequently proceeded to
parent body formation via gravitational collapse of dust clouds.

\section{Matrix processing fractions}

\subsection{Semarkona's chondrules and matrix}
\label{sec_Semarkona}

\begin{figure*}[t!]\begin{center}
\includegraphics[width=0.9\columnwidth]{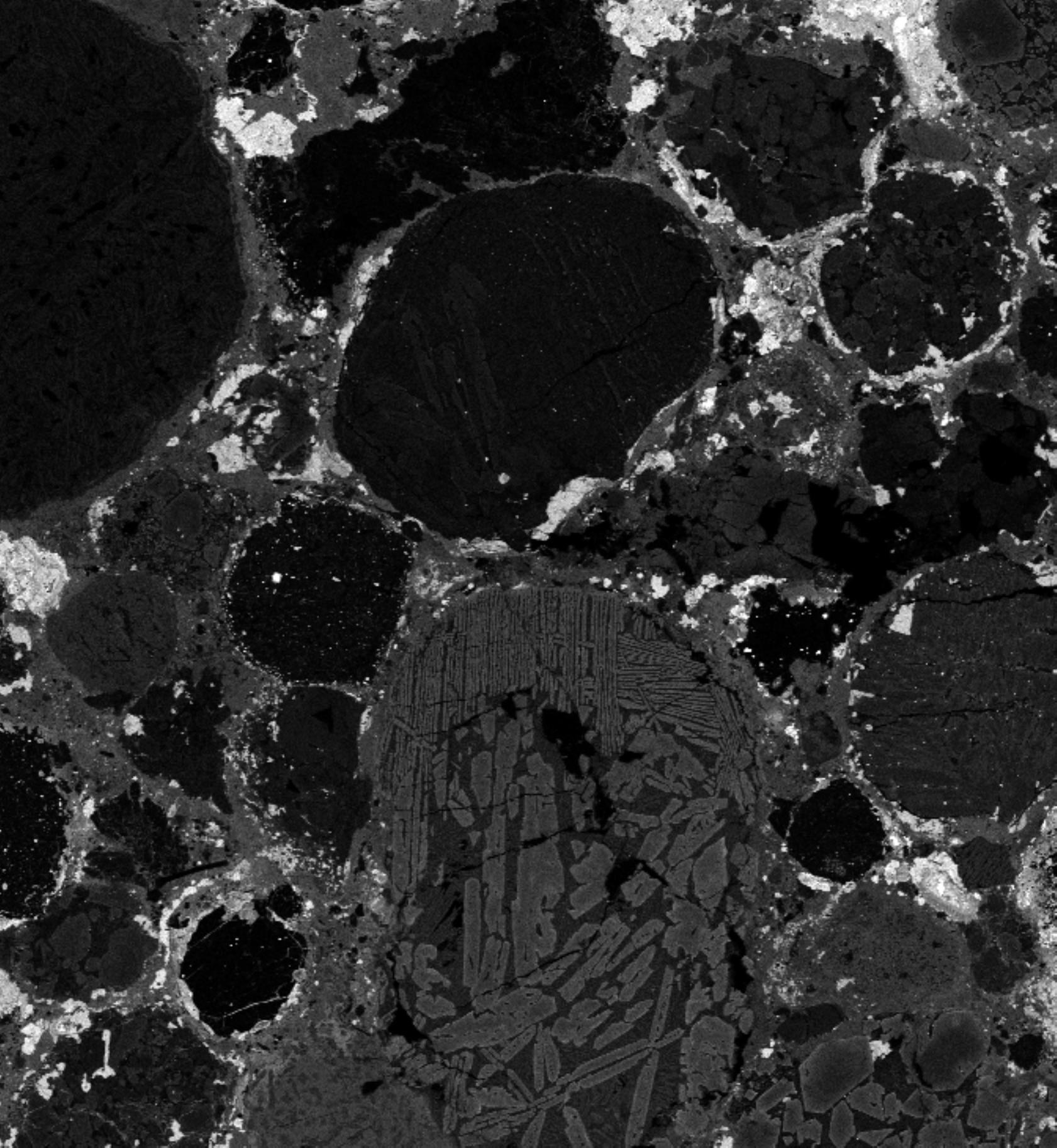}
\includegraphics[width=0.9\columnwidth]{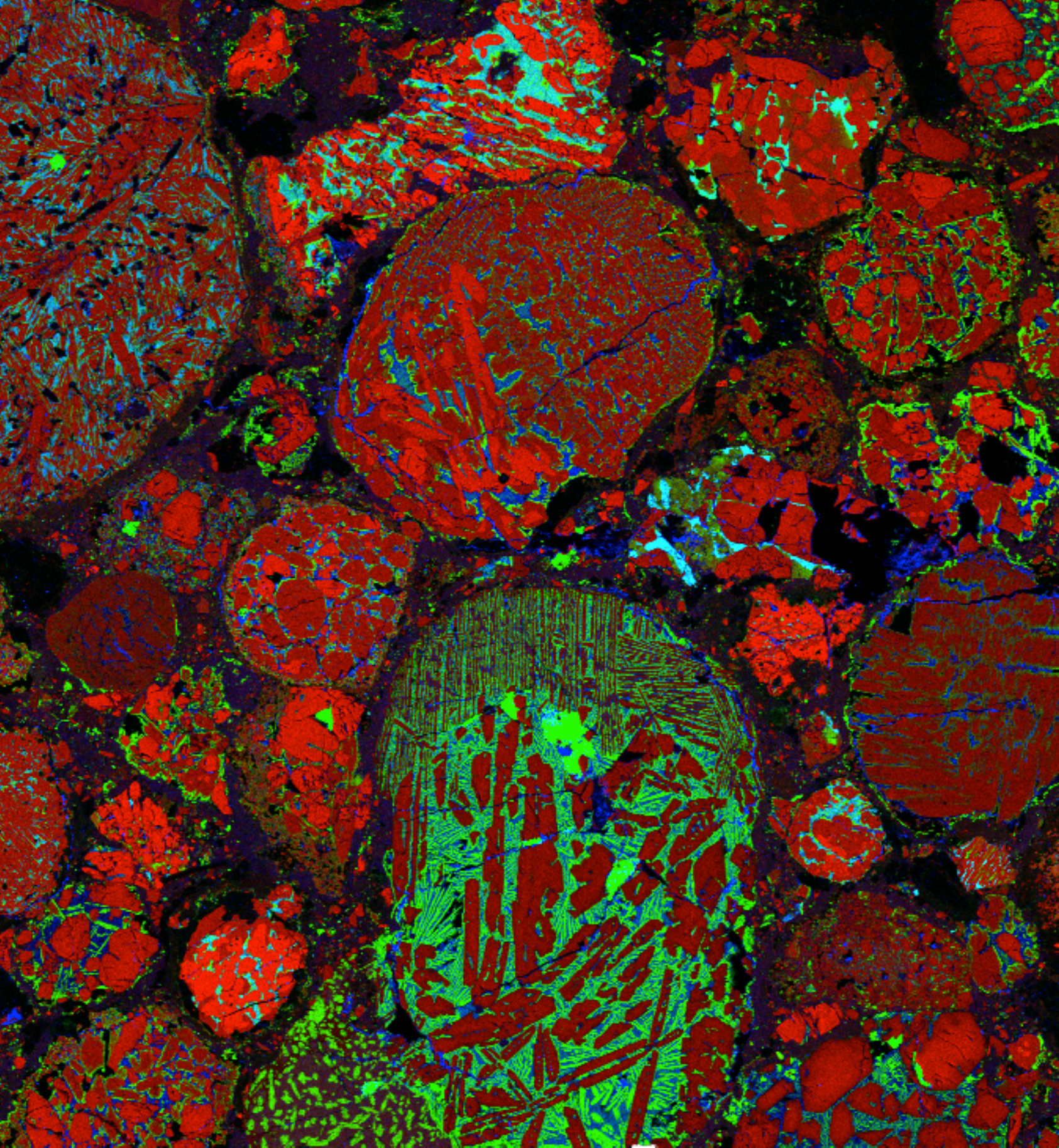}
\end{center}
\caption{
Slice through Semarkona (portion of sample AMNH 4128-t1-ps1A, $3.35 \times 3.65$~mm).
Left: Backscattered electron microscope.
Right: Mg (red)-Ca (green)-Al (blue). 
One can see many sub-mm sized chondrules, and in the bottom-center a multi-zoned chondrule
which was made by aggregating at least three chondrules.
\label{Sem_Slice} 
}
\end{figure*}

Semarkona is an LL3.0, shock stage 2 meteorite made up mostly of chondrules about $0.5$\,mm in diameter, plus about $\sim 27 \%$
fine grained matrix by surface area  \citep[see also \Fig{Sem_Slice}]{2005M&PS...40...87G, 2014LPI....45.1423L, Friedrich2014}.
In this paper we assume that the chondrules were made by melting
free-floating dust clumps, which requires ambient temperatures above $~1700$\,K \citep{1990Metic..25..309H}.
We use the term matrix to refer not only to the existing matrix material in Semarkona today, but also
any dust in the Solar Nebula which would today be classified as matrix (i.e.~not a chondrule) were it incorporated into Semarkona.
Our model considers a time span during which heating converts matrix into chondrules, so the fraction of material
labeled as matrix decreases over the interval we consider.

This free-floating dust was likely not pure pristine ISM material, presumably including
already thermally processed material such as relict grains \citep{2012M&PS...47.1176J}.
For simplicity we assume that most of the thermal processing experienced by
the non-chondrule portion of Semarkona was part of the chondrule forming process, and not a contaminant
from elsewhere/elsewhen.  In our framework this is a conservative hypothesis: if external heating is
significant, then the amount of not-chondrule forming heating allowed in \Eq{Eq_interval} is reduced, imposing
stricter constraints.

Interestingly, while Semarkona's chondrules have a size spread, there are almost
no $1$\,mm diameter chondrules in Semarkona, and no chondrules significantly larger than that \citep{2014LPI....45.1423L}.
This means that agglomerations
of many average chondrules were not themselves melted,
even though some of Semarkona's condrules show signs of several melting events interleaved with
collisional growth
\citep[see also \Fig{Sem_Slice}]{1996cpd..conf..119W,2013M&PS...48..445R}.

About half of 
Semarkona's matrix material was heated enough, above circa $800$\,K,
to release the P3 gas component from the (likely) pre-solar nano-diamonds
\citep{1994Metic..29..811H}, and other temperature probes suggest similar matrix temperatures
for chondrites \citep{1999Sci...285.1380B}
Of course, the heating that did occur may have happened after parent-body formation: more heavily altered meteorites show
much more matrix heating.
Nonetheless, even though these measurements provide only upper limits to the heating experienced
by matrix material before parent body assemblage, they still allow us to construct a simple model to estimate how much
of the ambient dust was melted into chondrules.
We define $c$ as the mass ratio of chondrules to all solids, $m$ as the mass ratio of all matrix material to all solids ($c+m=1$)
and $m_l$ as the mass ratio of non-heated (P3 not released) matrix material to all solids.  

\subsection{Thermal processing rates}
\label{therm_proc_rates}

\begin{figure}[t!]\begin{center}
\includegraphics[width=\columnwidth]{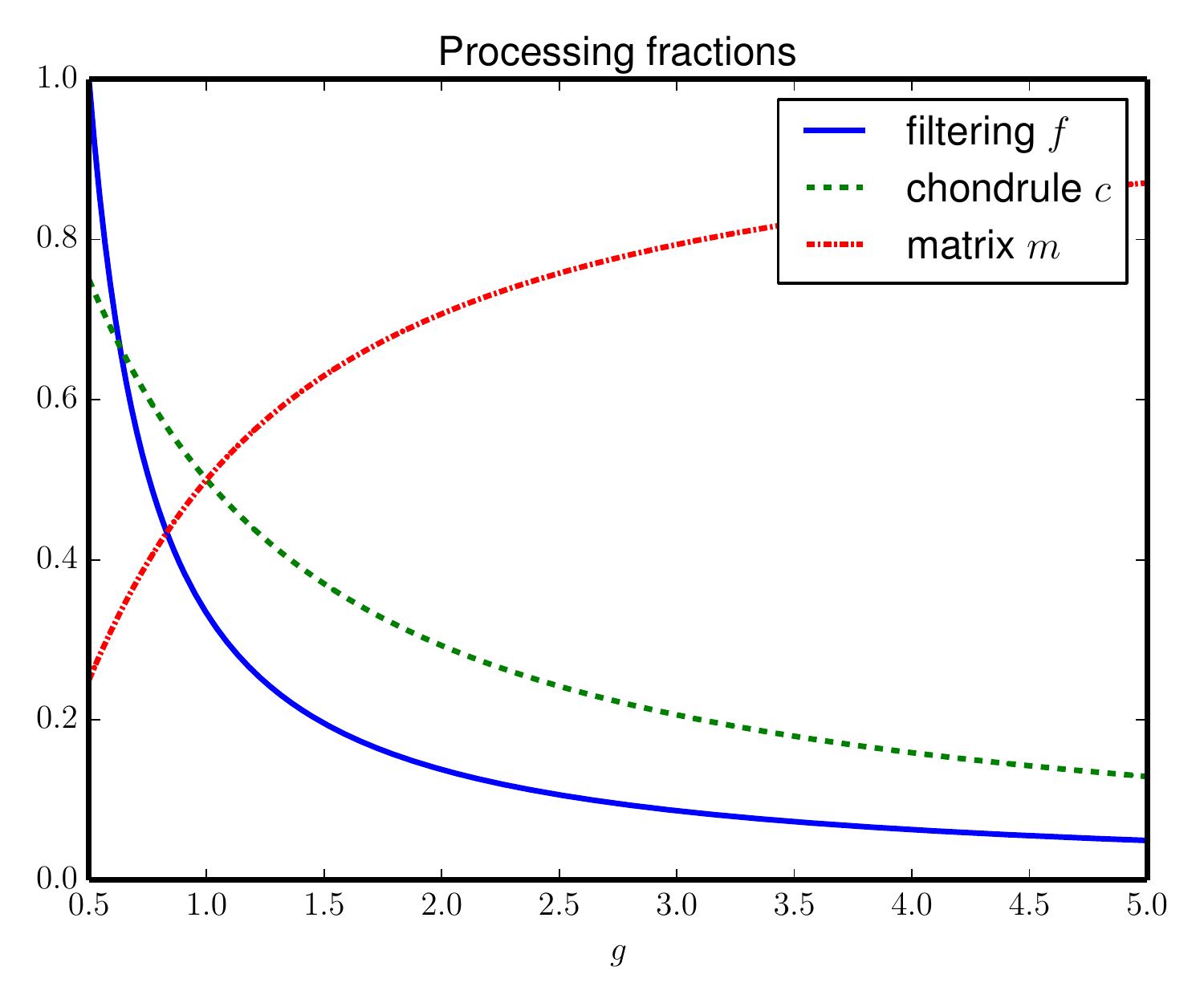}
\end{center}
\caption{
Filtering fraction $f$, chondrule mass fraction $c$, and matrix mass fraction $m$ required
to match Semarkona's chondrule fraction and matrix heating limits,
as a function of the heating parameter $g$.
\label{fig_filtering} 
}
\end{figure}

We assume that solids encounter regions hot enough to make
chondrules at a rate $k_c$.  These regions have warm sheathes, too cool to melt chondrules, but hot enough to
liberate the P3 component of the gas from the matrix, so every chondrule melting event
will also process additional matrix material without melting it.

We can parameterize the rate $k_h$ at which matrix loses the P3 component while remaining
matrix material in terms of the chondrule melting rate:
\EQ
k_h = g k_c.
\EN 
We will quantify our results in terms of the (poorly constrained) parameter $g$ 
which measures the rate at which dust is heated enough to release the P3 component, but
not to melt.  In many scenarios $g$ reduces to the ratio of the volume of the warm sheathes around
chondrule melting regions to the volume of those melting regions.

We also assume that the total system has a lifetime $t$, which we
decompose into $n$ intervals with $t=n \delta t$ and $n$ large enough that
$b=k_c \delta t$, with $b, gb \ll 1$. 
This allows us to estimate that, after a time $n \delta t$, the fraction of non-heated matrix  to total solids and 
the fraction of total matrix to total solids are
\EQ
m_l \simeq (1-[1+g]b)^n,
\EN
and
\EQ
m \simeq (1-b)^n,
\EN
respectively.  From the pre-solar grain noble gas measurements we know that
\EQ
\frac{m_l}{m} = \frac{(1-[1+g]b)^n}{(1-b)^n} \gtrsim \frac 12. \label{Eq_interval}
\EN
Taking the logarithm of \Eq{Eq_interval} and using $b, gb \ll 1$ we find
\EQ
n([-b-gb]-[-b]) \gtrsim - \ln 2,
\EN
so 
\EQ
bn \lesssim \ln 2^{1/g}.
\EN
Finally, we find that the fraction of material not turned into chondrules is
\EQ
m \simeq \exp \left( \ln \left[(1-b)^n\right] \right) \gtrsim \frac{1}{2^{1/g}}.
\EN
We assume equality henceforth both
for simplicity, and as an lower limit for the strength of the constraint.

\subsection{Need for filtering}

However, Semarkona is $\sim 75\%$ chondrule, so the fraction of non-chondrule to chondrule material in Semarkona is $1/3$.
If the planetesimal formation process makes use of all
the chondrules and a fraction $f$ of the matrix material, then
\EQ
\frac{mf}{c} = \frac{ 2^{-1/g}f}{1-2^{-1/g}} =\frac 13.
\EN
Solving for the filtering fraction $f$, we arrive at
\EQ
f= \frac 13 \left(2^{1/g}-1\right),
\EN
plotted in \Fig{fig_filtering}.
Note that if $g<0.5$, then chondrules make up more than $75\%$ of the solids, so they, not matrix, need to be filtered out.

While the parameter $g$ is as yet unstudied, the large difference between the chondrule melting temperature ($\sim 1700$\,K)
and the matrix heating temperature ($\sim 800$\,K) suggests that hot sheathes around chondrule melting zones should be large.
\cite{2013ApJ...776..101B}, a study of planetesimal bow shocks as a chondrule formation mechanism, 
did not quantify the parameter $g$, but its figures suggest that $800$\,K is reached for impact parameters at least twice
that required for melting chondrules, or $g>3$.  The initial conditions used in \cite{2014ApJ...791...62M}, a study
of magnetic heating, are above $800$\,K,
which limits its ability to constrain $g$ when interpreted as a model for a full disk as opposed to surface layers.
However, that work also suggests large warm regions, i.e.~a respectable $g$.
Further, while there are quite a few mechanisms proposed to reach temperatures above $1700$\,K, those mechanisms can
also fail, resulting in heating episodes that never achieve chondrule melting temperatures raising $g$ even higher \citep{2013ApJ...767L...2M}.
The authors think that $g>1$ is a quite conservative estimate;
and even that value requires $f<1/3$, which means that more than $2/3$ of the matrix material must have been filtered out.

\subsection{Complementarity}

\begin{figure}[t!]\begin{center}
\includegraphics[width=\columnwidth]{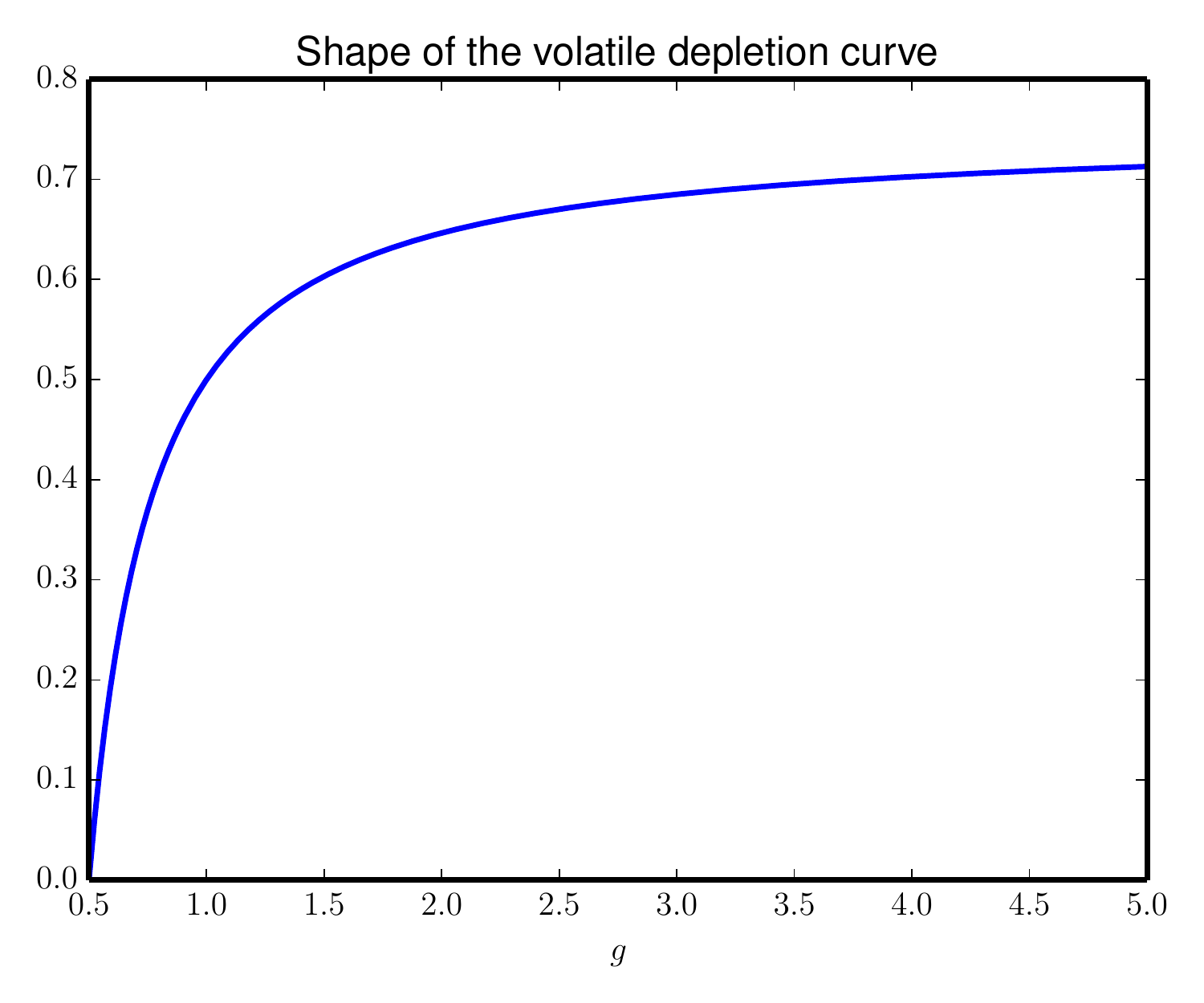}
\end{center}
\caption{
Shape of the volatile depletion curve (\Eq{compl}  with $e=1$). 
\label{fig_compl} 
}
\end{figure}

If the matrix and chondrules originally had different abundances from each other,
then filtering would have altered the abundances of the assembled
whole as compared to the mean abundances of the Solar Nebula's total solids.
In particular, chondrule melting presumably evaporated volatiles which (in part) recondensed onto the matrix,
some of which survived as matrix up to the parent body assemblage stage.  If
some of that surviving matrix was filtered out,
 the resulting chondrite would be volatile depleted.

As an LL chondrite, Semarkona is noteworthy for
its low iron (and siderophile) abundances, a topic beyond the scope of this paper.  However (once
normalized to magnesium) Semarkona is
approximately Solar in lithophile abundances down to the moderately volatile elements such as sodium and potassium
\citep{2006mess.book...19W}, even though its matrix and chondrules have moderately different abundances from one
another \citep{2014LPI....45.1423L}.
This anti-correlation between matrix and chondrules is known as ``complementarity'' \citep{2010E&PSL.294...85H}.

If a given element evaporated during the chondrule melting process,
a significant fraction of the evaporated element would have recondensed onto the newly formed chondrules
due to physical proximity.  Nonetheless, we can approximate that
a fraction $e$ of the element would have both evaporated from the chondrules and recondensed on the matrix, which would
otherwise have had identical abundances of that element.  This results in the chondrules being depleted in the element by
a fraction $e$, the matrix being enriched in the element by a fraction $ec/m$ and
the final bulk abundances depleted by a fraction
\EQ
e\frac{c-fc}{c+fm}, \label{compl}
\EN
the shape of which is plotted in \Fig{fig_compl}.

The depletion
curve is relatively insensitive to $g$ once $g>1$,
so strong depletion is expected if $e$ is close to unity.
However that would mean that there was very little recondensation onto the chondrules.
Very little volatile depletion is seen in Semarkona, which constrains $e \ll 1$, and means that the chondrule
melting region must have been large enough for most of the recondensation to occur before mixing
with the surrounding, matrix grain filled gas.  This is conceptually similar to but weaker than the constraints
derived from Semarkona's chondrules' sodium content
put forth in \cite{2008Sci...320.1617A} and expanded on since \citep{Hewins12}.

We also note that complementarity  and filtering can only be reconciled if parent body formation occurred soon after chondrule melting:
significant filtering at the drag parameters we will discuss in \Sec{sec_collective_behavior} 
means that matrix and chondrules drifted radially at different rates.  Dust grains with $St=10^{-2}$ drift radially at about $1$\,m\,s$^{-1}$, or about $5000$\,yr\,AU$^{-1}$
\citep{1977MNRAS.180...57W}.  This would erase complementarity on a kyr time scale.

\section{Assembling boulders and planetesimals}
\subsection{Stokes numbers}

The dynamics of small dust grains in protoplanetary disks are controlled by their interactions with the ambient gas, which occurs through drag:
\EQ
\frac {\partial \vv}{\partial t} = -\frac{\vv-\uu}{\tau}, \label{dvdt}
\EN
where $\vv$ is the dust grain velocity, $\uu$ the local gas velocity and $\tau$ the frictional stopping time.  We non-dimensionalize $\tau$
by defining the Stokes number $St$ as
\EQ
St \equiv \tau \Omega,
\EN
where $\Omega$ is the orbital frequency.

\Eq{dvdt} implies that dust dynamics depends solely on the drag parameter.
For filtering to have occurred, the ambient matrix grains and the chondrules (or chondrule assemblies)
that went into Semarkona had to have had different Stokes numbers.  
The chondrule melting process itself can alter $St$ \citep{2014Icar..237...84H}.
Dust Stokes numbers at the midplane
of a disk are (assuming Epstein drag, appropriate for naked Semarkona chondrules):
\EQ
St = \sqrt{2 \pi} \frac{a \rho_s}{\Sigma_g},
\label{ST}
\EN
where $a$ is the dust grain radius, $\rho_s$ the dust solid density and $\Sigma_g$ the gas disk surface density.
The chondrule melting process eliminates the matrix's porosity, and so increases the grains' densities $\rho_s$.
This in turn increases the Stokes numbers of the chondrules
as compared to the Stokes numbers of the matrix precursors.

\subsection{Dust gravitational instabilities}
\label{sec_dust_grav}

At some point the cloud of chondrules which would make up Semarkona must have become gravitationally unstable, allowing the formation
of Semarkona's parent body.
Using the observed Semarkona chondrule mean radius $a=0.25$\,mm, and a 
chondrule solid density of about $\rho_s = 3$ g\,cm$^{-3}$, \Eq{ST} implies that a naked chondrule has
\EQ
St \simeq 2 \times 10^{-3} \left(\frac{100 \text{\,g\,cm}^2}{\Sigma_g}\right). \label{St_c_100}
\EN
Note that $\Sigma_g=100$\,g\,cm$^{-2}$ is well below the Hayashi Minimum Mass Solar Nebula (MMSN) value of $\Sigma_g = 430$\,g\,cm$^{-2}$ at $R=2.5$AU,
which in turn is far smaller than the more recent Desch MMSN value \citep{1981PThPS..70...35H,2007ApJ...671..878D}.
Accordingly, \Eq{St_c_100}'s value for chondrule Stokes
numbers is a significant overestimate.

Even the largest, $\sim 1$\,mm diameter, chondrules in Semarkona have midplane $St \ll 10^{-2}$ for reasonable
gas surface densities.  We also note that at most half the solids could have been melted into chondrules ($g\ge 1$ implies
$c<0.5$, see \Fig{fig_filtering}).  This means that
the expected chondrule-to-gas surface density ratio is less than half the $0.5\%$ rock-to-gas mass ratio expected
for the overall Solar Nebula, itself less than the $1.5\%$ ice+rock-to-gas mass ratio \citep{2003ApJ...591.1220L}.

Clouds of very low $St$ dust can be self-gravitating, but achieving that is difficult.  If there is any background turbulence,
it will stir the dust, preventing the onset of gravitational instabilities \citep{1980Icar...44..172W}.
Even if there is no background turbulence, the
dust will need to have settled into a very thin, Kelvin-Helmholz unstable layer.  Our current understanding
is that if dust is nearly perfectly coupled to the gas ($St \ll 1$), it 
can become gravitationally unstable only for significantly, $>4 \times$, super-Solar Nebular solid-to-gas mass ratios
\citep{2006Icar..181..572W,2010ApJ...725.1938L}.  However, because most of the solids were in the free floating matrix,
 Semarkona's chondrules have a much lower chondrule-to-gas mass ratio: that ratio
 for the chondrule population is $c/3<1/6$ for $g>1$; where the factor of $3$ converts between rock+ice
 to just rock as noted above.  Direct formation
 of planetesimals through unaugmented
 gravitational instabilities of naked Semarkona chondrules is therefore ruled out.

\subsection{Collective behavior}
\label{sec_collective_behavior}

The above constraints can be alleviated by collective behavior such as the Streaming Instability (SI).
In the $St < 1$ regime however, collective behavior between dust grains that lead to gravitational interactions still
require significantly higher dust-to-gas surface density ratios,
closer to $3\%$ (two times the expected value for the Solar Nebula), and,
further, those interactions only set in once the dust grains grow large enough that their Stokes numbers are
$St > 10^{-2}$ \citep{2009ApJ...704L..75J,2010ApJ...722.1437B}.  In the case of naked Semarkona chondrules, that would mean a gas surface density
less than $20$\,g\,cm$^{-2}$, less than $5\%$ of the Hayashi MMSN value, combined with an, at minimum, order of magnitude enhancement in the
local chondrule-to-gas surface density ratio above values suggested by \cite{2003ApJ...591.1220L}.

From \Eq{ST}, we can see that to reach the minimum $St \gtrsim 10^{-2}$ required for the SI to act, 
non-porous objects with $\rho_s=3$\,g\,cm$^{-3}$ in a MMSN with $\Sigma_g=430$\,g\,cm$^{-2}$ would need to
have a radius
\EQ
a \gtrsim \frac{ 10^{-2} \Sigma_g}{\sqrt{2 \pi} \rho_s} \simeq 0.6\,\text{cm},
\EN
more than 20 times larger than (and more than $10^4$ times as massive as) Semarkona's $a=0.25$\,mm chondrules.
Allowing for $\sim 50\%$ porosity, the radius
increases to above $1$\,cm.
Distinct clusters of chondrules, cluster chondrites, much closer to the size required by the SI (a significant fraction of $1$\,cm) are
found in the meteoritical record, although their abundance in Semarkona is as yet unmeasured \citep{2012M&PS...47.2193M}.
While not a dominant feature of their host chondrites, these clasts may represent the agglomerations required
to proceed to the gravitational collapse of dust clouds in the Solar Nebula.

We conclude therefore that, even considering collective behavior such as the SI,
naked Semarkona chondrules could not have proceeded to gravitational collapse into a parent body without
significant intervening collisional growth: current dust dynamical theory does not otherwise
allow for parent body formation.
This also means that any changes in $St$ that resulted from melting cannot explain
the filtering out of matrix material because those were not the $St$ values of the dust grains when gravity took over.

Significantly, the agglomerations that must resulted from collisional growth are not dominant in the meteoritical record,
and the results of melting such agglomerations are not present.  It follows that the chondrule melting location
and the chondrule-agglomeration assemblage location must have been distinct, either in space or time.

\section{Collision resilience}
\label{Sec_Col_Rel}

The sizes, and hence Stokes numbers, reached by collisionally growing dust grains are controlled by bouncing and fragmentation.
Dust grain collision speeds are determined by their Stokes numbers, with 
radial drift collision speeds scaling with $St$ \citep{1977MNRAS.180...57W}.
Turbulent induced collision speed scales as
\EQ
v_c = \beta \sqrt{\alpha St} c_s \label{vc}
\EN
where $\alpha$ is Shakura-Sunyaev $\alpha$ parameter, here used to parameterize
the turbulent diffusivity \citep{1973A&A....24..337S} and $\beta$ a factor of order unity.  \Eq{vc} can be obtained on dimensional grounds
\citep{1980A&A....85..316V} and numerical simulations have confirmed that $\beta \lesssim 1$
\citep{2012MNRAS.426..784H,2013MNRAS.432.1274H, 2013ApJ...776...12P,2014ApJ...791...48P}.
Combining these scalings with \Eq{ST} we can see that dust growth results in faster collisions.

When the characteristic collision speeds breach a threshold bouncing  speed $v_b$, dust-dust encounters start resulting in growth-neutral bouncing events \citep{2010A&A...513A..57Z}.  This slows collisional growth, but the low velocity
tail of the collision velocity distribution still allows coagulation \citep{2012A&A...544L..16W}.
Laboratory studies have shown that $v_b$ is on the order of $1$\,mm\,s$^{-1}$ for the
$\sim 10^{-4}$\,g chondrules and chondrule precursors we consider \citep{2010A&A...513A..56G}.
At an even higher collision speed
threshold, $v_f \sim 100$\,cm\,s$^{-1}$, dust grains fragment,
placing an effective cap on the dust grain $St$ values, and hence their sizes.  A pile-up in dust sizes with $St$ values that concentrate the
collision speeds above $v_b$ but below $v_f$ has been suggested as a cause of the narrowness of the chondrule size distribution
\citep{2014Icar..232..176J}.

Chondrule agglomerations must have grown collisionally in the Solar Nebula because Semarkona's chondrules 
were too small to directly form parent bodies
 (\Sec{sec_collective_behavior}).
The size distribution of these agglomerations was
controlled by $v_b$ and $v_f$.  However, matrix grains also grew collisionally, with their own $v_b$ and $v_f$.
To achieve filtering, the final, collisionally determined, $St$ distribution of the matrix must have been different from that of the chondrule
agglomerations. 

Collision velocities themselves are only a function of $St$, so if chondrule assemblages interacted with each other
with the same critical $v_b$ and $v_f$ velocities as the matrix, then both the chondrule assemblages and the matrix would
have grown to the same $St$.  If that were the case, matrix material could not have been filtered out at the planetesimal formation
stage because the matrix and the chondrule assemblages would have obeyed identical dynamical equations.
It follows that the fragmentation speed
$v_f$ (and likely the bouncing speed $v_b$) of chondrule assemblages
must have been significantly higher than that of the matrix. Further laboratory experiments will be required to quantify 
this difference.

\section{Chondrule melting location}
\label{chond_melt_loc}

Given that large chondrule agglomerations must have formed
in the Solar Nebula (\Sec{sec_collective_behavior}), 
it is surprising both that no melted high $St$ agglomerations are seen in Semarkona, and
that even unmelted distinct assemblages of chondrules are not dominant.
Indeed, while there is evidence of small, mm-sized
melted chondrule agglomerations or chondrule-matrix agglomerations
\citep[see also \Fig{Sem_Slice}]{2013M&PS...48..445R}, Semarkona's chondrules have a conspicuously narrow
size range \citep{2014LPI....45.1423L}.  This implies that, once melted, chondrules must have exited the melting zone
on short timescales, before they could grow collisionally and be remelted.
Semarkona chondrules were however limited in their ability to move large distances radially: their very low
$St \ll 1$ means that they were well coupled to the gas, and so their radial motion was tied to the accretion flow of the disk.

Observations of extrasolar protostars suggest that the quasi-steady state accretion rate through the Solar Nebula was between
$10^{-10}$ and $10^{-7.5}\,M_{\odot}/$yr \citep{2013ApJ...767..112I}.
This rate can be surpassed only briefly, for example during brief episodic accretion events such
as FU Orionis type events \citep{1996ARA&A..34..207H}.
For an Hayashi MMSN at $R=2.5$\,AU, the upper end of
that accretion rate range implies an accretion velocity  $v_r=20$\,cm\,s$^{-1}$ and a characteristic radial transport time
$R/v_r = 60$\,kyr.  Those values are consistent with a viscous alpha-disk with $10^{-3} < \alpha < 10^{-2}$
\citep{1973A&A....24..337S}.  This means that removing chondrules from even a narrow radially demarcated melting zone would have taken
too long to avoid melting large chondrule agglomerations.

Vertical settling is a more plausible way to have removed the chondrules from the melting zone.  The vertical distribution of dust grains
is controlled by the competition between gravity pulling the grains to the midplane and turbulence stirring the grains upwards.
Dust, especially at the sub-mm size scale considered here, is expected to be very porous
\citep{2007A&A...461..215O,2008ARA&A..46...21B},
so non-porous chondrules have higher $St$ than their (porous) precursor matrix grains.
If the melting was restricted to surface layers, then the higher $St$ chondrule would have settled out of the
melting layer on sub-century time scales \citep{2014Icar..237...84H}.  Subsequently formed chondrule agglomerations, with even higher $St$ values, would have been unable to reenter the melting layers.
Settling does not further complicate complementarity because it requires that the matrix material is vertically well mixed by turbulence.

Further, this scenario also helps explain the fact that Semarkona's chondrules are both small and have a narrow size distribution.  With $St$
numbers well below $St=10^{-3}$, the precursors to Semarkona's chondrules would naively be expected to collide slowly
enough to grow for all but the most violent turbulent stirring (see \Sec{Sec_Col_Rel}).
However, dust grains of a given midplane $St$ are well mixed with the gas up to a height
\EQ
H_d =\sqrt{2 \ln \left(1+\frac{\alpha}{St}\right)} H_g
\EN
where $H_g$ is the gas scale height and $\alpha$ the Shakura-Sunyaev $\alpha$ parameter used to measure the turbulent
diffusivity of the dust as in \Sec{Sec_Col_Rel}
\citep{1995Icar..114..237D, 2002ApJ...581.1344T}.  This behavior is well fit by numerical simulations of turbulent disks \citep{2006A&A...452..751F,
2006MNRAS.373.1633C}.
This means that only matrix grains with small
enough $St$ values could be lofted into the melting zone and provides a complement to the bouncing barrier in limiting
the size distribution of chondrules \citep{2014Icar..232..176J}.

These constraints remain even when the ability of large dust grains to survive the melting process is
considered.  In the planetesimal bow shock model \citep{1998M&PS...33...97H}, dust grains find themselves moving through gas
at high velocity.  \cite{2002ApJ...564L..57S} considered the competition between ram pressure and surface tension for chondrule melts, and their
results imply that $a \sim 1$\,cm grains could survive the shocks invoked by, e.g., \cite{2012ApJ...752...27M}.  That size is well above the observed
chondrule size for Semarkona, so the fragmentation of melted chondrules by shocks is not an explanation for the lack of large chondrules.

\section{Discussion and conclusions}

We have identified three important constraints on the formation of Semarkona (and other chondrites).  Firstly, laboratory
studies have shown that Semarkona is a mixture of high temperature chondrules and cold matrix.
Any reasonable chondrule melting
mechanism will also heat a significant amount of matrix material enough to have experimental consequences,
so this allows us to conclude that the chondrule-to-matrix ratio in the Solar Nebula must have been low.
This is the case even
though the experimental limits on the matrix's thermal history are weak, and quite possibly dominated by parent body processing
rather than nebular processing \citep{1994Metic..29..811H}.  As a result, for our uses the constraints are upper limits on the
chondule-to-matrix ratio in the Solar Nebula.
The large amount of matrix material that was never heated to high temperatures (above $800$\,K)
also rules out large scale spiral waves \citep{2005ApJ...621L.137B} as a chondrule melting mechanism
for Semarkona because such waves would spare nothing.

Secondly, analytical and numerical studies of dust concentration have shown that naked Semarkona chondrules were too small
to have directly proceeded to parent body formation.  This problem is exacerbated by the modest chondrule melting efficiency mentioned
above.  It immediately follows that large chondrule agglomerations must have been made, where
``large'' means more than $20$ average chondrule radii in size \citep[so larger than the agglomerations
considered in][]{1996cpd..conf..119W}.
Finally, Semarkona's chondrule-to-matrix ratio is much higher than the small nebular chondrule-to-matrix
implied by the requirement that the matrix have remained cold.  It follows that
some form of filtering out of matrix material during parent body formation occurred.

From these constraints we can infer two significant consequences.  Firstly, the results of melting large chondrule
agglomerations is not seen in the meteoritical record \citep{Friedrich2014}, even though the agglomerations must have formed.  We can therefore conclude
that chondrules exited the melting region rapidly, before the agglomerations reached their final size.  Due to the difficulty in moving
objects the size of Semarkona's chondrules radially in the Solar Nebula, we suggest that this motion was probably vertical settling to the midplane,
assisted by the reduction in the drag parameter associated with melting a porous dust grain into a non-porous chondrule.
This disfavors planetesimal bow shocks
\citep{1998M&PS...33...97H,2012ApJ...752...27M} as the chondrule melting process for Semarkona
because a significant number of the planetesimals producing those shocks must have been on low inclination orbits, resulting in midplane heating

On the other hand, magnetic heating mechanisms such as the short-circuit instability
are well suited to surface layer heating  \citep{2012ApJ...761...58H,2013ApJ...767L...2M}.
Protoplanetary disks are expected to have radially extended magnetically dead-zones where the ionization fraction is too low to
support magnetic fields.  However, these dead-zones are bounded above and below
by magnetically active layers where non-thermal ionization is sufficient for
magnetic fields to couple to the gas \citep{1996ApJ...457..355G}.
This would naturally result in magnetically-mediated chondrule formation mechanisms
having operated in upper layers of the Solar Nebula but not at the midplane.

Secondly, because matrix material had to have been filtered out at some stage of planetesimal formation, we conclude that
the matrix dust must have had different drag parameters than the chondrule agglomerations.  Theory predicts that collisional coagulation continues
until dust grains grow large enough that their collisions result in bouncing or fragmentation, so we suggest that
chondrule agglomerations were significantly more collision-resilient than chondrule-free matrix grains.  
This is not surprising, as chondrules, solid sub-mm objects,
could develop coatings of fine dust  \citep{2013EP&S...65.1159F}, which
have been shown to dramatically increase the ability of chondrule analogs to stick
\citep{2012Icar..218..701B}.  Fractal matrix grains, made of sub-micron 
constituents, would not have been able to
accrete such a coating.
Further, chondrules could
act as the weights of a bola, wrapping up chondrule agglomerations into low porosity, high contact area objects.

This would be convenient because experiments performed for matrix-like material have found
a fragmentation speed $v_f \sim 100$\,cm\,s$^{-1}$ \citep{2010A&A...513A..56G}.
Those fragmentation speeds are difficult to reconcile with models of collective behavior and dust gravitational instabilities
because they predict that dust grains large enough to trigger the collective behavior would be destroyed in dust-dust collisions
for the bulk of protoplanetary disk model parameter space.
Higher $v_f$ chondrule agglomerations therefore provide an easier route to self-gravitating dust.

Our picture of surface layer chondrule melting is in tension with the picture put forward in \cite{2008Sci...320.1617A} to
explain the sodium found in Semarkona's chondrules. They found that the distribution of sodium
within Semarkona's chondrules implies that they were melted in a dust cloud dense enough be gravitationally
unstable, while high altitude melting would suggest relatively low dust densities.  In this paper we have shown that
concentrating naked Semarkona chondrules to a gravitationally unstable
density is impossible under our current theoretical understanding.  The problem is even worse for matrix
precursors  with even lower Stokes numbers.  Further, if the self-gravitating dust cloud proposed in \cite{2008Sci...320.1617A} was heated enough to melt chondrules,
there would not be any surviving matrix material.  This is a particular problem because 
a not-insubstantial amount of new matrix (not just cold matrix) would need to have been mixed in before the self-gravitating
dust cloud collapsed to fit Semarkona's chondrule-to-matrix ratio.

Other scenarios, such as chondrule melting in impact plumes
which locally enhance the sodium partial pressure \citep{Fedkin13}, have been proposed to explain sodium measurement.  However, many chondrules do not
possess abundant sodium \citep{2006mess.book...19W}, so some chondrule formation did result in the loss of volatiles.
The diversity of chondrules means that multiple formation scenarios likely occurred; but if so, Semarkona's
chondrules could not have spent much time in a region where the volatile-depleting chondrule melting happened, or they would have been remelted
and sodium depleted.  This reinforces the argument that the chondrules must have exited the melting region rapidly, but the inference that
melting occurred in surface layers can only be drawn if sodium-preserving melting scenarios can operate there.

While this analysis has focused on Semarkona, it also applies more generally.  Semarkona's chondrules are large for Ordinary Chondrites
(OCs), and larger than chondrules in many Carbonaceous Chondrites (CCs) \citep{2006mess.book...19W}.
Accordingly, the observation that chondrules are too small to directly proceed to gravitational collapse or collective behavior such
as the Streaming Instability is general.  It follows that the need for forming
large agglomerations is also general, although for in the case of CCs, with lower chondrule volume fractions, these agglomerations need not have been chondrule dominated.  
The result of melting these agglomerations is also not seen in the general meteoritical record, so the requirement of
getting the chondrules out of their melting zone rapidly, before they could be incorporated into agglomerations, also remains.

While all OCs are dominated by chondrules, suggesting that filtering was needed, the chondrule mass fraction is lower in the CCs.  For CCs, the need for filtering out matrix material depends much
more significantly on the limits that can be placed on the matrix's thermal history.  Unfortunately, the matrix-material temperature history limits
are poor, and convolved with parent body alteration, but the evidence suggests that a large fraction of the matrix material stayed quite
cool across chondrites \citep{2011M&PS...46..252A}.
However, we hope that, having shown that those temperature histories are an important
clue to the nature of the Solar Nebula, we will see more experimental results constraining the maximum
Solar Nebula temperature experienced
by the matrix across the chondrite classifications.  Further, we also hope that future numerical studies of chondrule heating
will also consider the effect that lower temperatures have on the dust, and quantify the rate at which matrix material is heated.

\section*{Acknowledgments}

This research has made use of the National Aeronautics and Space AdministrationÕs Astrophysics Data System Bibliographic Services.
The work was supported by National Science Foundation, Cyberenabled Discovery Initiative grant AST08-35734, 
AAG grant AST10-09802, NASA OSS grant NNX14AJ56G and a Kalbfleisch Fellowship from the American Museum of Natural History.

\bibliography{semarkona}

\begin{thebibliography}{57}
\expandafter\ifx\csname natexlab\endcsname\relax\def\natexlab#1{#1}\fi
\providecommand{\url}[1]{\texttt{#1}}
\providecommand{\href}[2]{#2}
\providecommand{\path}[1]{#1}
\providecommand{\DOIprefix}{doi:}
\providecommand{\ArXivprefix}{arXiv:}
\providecommand{\URLprefix}{URL: }
\providecommand{\Pubmedprefix}{pmid:}
\providecommand{\doi}[1]{\href{http://dx.doi.org/#1}{\path{#1}}}
\providecommand{\Pubmed}[1]{\href{pmid:#1}{\path{#1}}}
\providecommand{\bibinfo}[2]{#2}
\ifx\xfnm\relax \def\xfnm[#1]{\unskip,\space#1}\fi
\bibitem[{{Abreu} \& {Brearley}(2011)}]{2011M&PS...46..252A}
\bibinfo{author}{{Abreu}, N.~M.}, \& \bibinfo{author}{{Brearley}, A.~J.}
  (\bibinfo{year}{2011}).
\newblock \bibinfo{title}{{Deciphering the nebular and asteroidal record of
  silicates and organic material in the matrix of the reduced CV3 chondrite
  Vigarano}}.
\newblock {\it \bibinfo{journal}{Meteoritics and Planetary Science}\/},  {\it
  \bibinfo{volume}{46}\/}, \bibinfo{pages}{252--274}.
  \DOIprefix\doi{10.1111/j.1945-5100.2010.01149.x}.
\bibitem[{{Alexander} et~al.(2008){Alexander}, {Grossman}, {Ebel} \&
  {Ciesla}}]{2008Sci...320.1617A}
\bibinfo{author}{{Alexander}, C.~M.~O.~.}, \bibinfo{author}{{Grossman}, J.~N.},
  \bibinfo{author}{{Ebel}, D.~S.}, \& \bibinfo{author}{{Ciesla}, F.~J.}
  (\bibinfo{year}{2008}).
\newblock \bibinfo{title}{{The Formation Conditions of Chondrules and
  Chondrites}}.
\newblock {\it \bibinfo{journal}{Science}\/},  {\it \bibinfo{volume}{320}\/},
  \bibinfo{pages}{1617--}. \DOIprefix\doi{10.1126/science.1156561}.
\bibitem[{{Bai} \& {Stone}(2010)}]{2010ApJ...722.1437B}
\bibinfo{author}{{Bai}, X.-N.}, \& \bibinfo{author}{{Stone}, J.~M.}
  (\bibinfo{year}{2010}).
\newblock \bibinfo{title}{{Dynamics of Solids in the Midplane of Protoplanetary
  Disks: Implications for Planetesimal Formation}}.
\newblock {\it \bibinfo{journal}{\apj}\/},  {\it \bibinfo{volume}{722}\/},
  \bibinfo{pages}{1437--1459}. \DOIprefix\doi{10.1088/0004-637X/722/2/1437}.
  \href{http://arxiv.org/abs/1005.4982}{\tt arXiv:1005.4982}.
\bibitem[{{Beitz} et~al.(2012){Beitz}, {G{\"u}ttler}, {Weidling} \&
  {Blum}}]{2012Icar..218..701B}
\bibinfo{author}{{Beitz}, E.}, \bibinfo{author}{{G{\"u}ttler}, C.},
  \bibinfo{author}{{Weidling}, R.}, \& \bibinfo{author}{{Blum}, J.}
  (\bibinfo{year}{2012}).
\newblock \bibinfo{title}{{Free collisions in a microgravity many-particle
  experiment - II: The collision dynamics of dust-coated chondrules}}.
\newblock {\it \bibinfo{journal}{\icarus}\/},  {\it \bibinfo{volume}{218}\/},
  \bibinfo{pages}{701--706}. \DOIprefix\doi{10.1016/j.icarus.2011.11.036}.
  \href{http://arxiv.org/abs/1105.3897}{\tt arXiv:1105.3897}.
\bibitem[{{Blum} \& {Wurm}(2008)}]{2008ARA&A..46...21B}
\bibinfo{author}{{Blum}, J.}, \& \bibinfo{author}{{Wurm}, G.}
  (\bibinfo{year}{2008}).
\newblock \bibinfo{title}{{The Growth Mechanisms of Macroscopic Bodies in
  Protoplanetary Disks}}.
\newblock {\it \bibinfo{journal}{\araa}\/},  {\it \bibinfo{volume}{46}\/},
  \bibinfo{pages}{21--56}.
  \DOIprefix\doi{10.1146/annurev.astro.46.060407.145152}.
\bibitem[{{Boley} et~al.(2013){Boley}, {Morris} \&
  {Desch}}]{2013ApJ...776..101B}
\bibinfo{author}{{Boley}, A.~C.}, \bibinfo{author}{{Morris}, M.~A.}, \&
  \bibinfo{author}{{Desch}, S.~J.} (\bibinfo{year}{2013}).
\newblock \bibinfo{title}{{High-temperature Processing of Solids through Solar
  Nebular Bow Shocks: 3D Radiation Hydrodynamics Simulations with Particles}}.
\newblock {\it \bibinfo{journal}{\apj}\/},  {\it \bibinfo{volume}{776}\/},
  \bibinfo{pages}{101}. \DOIprefix\doi{10.1088/0004-637X/776/2/101}.
  \href{http://arxiv.org/abs/1308.2968}{\tt arXiv:1308.2968}.
\bibitem[{{Boss} \& {Durisen}(2005)}]{2005ApJ...621L.137B}
\bibinfo{author}{{Boss}, A.~P.}, \& \bibinfo{author}{{Durisen}, R.~H.}
  (\bibinfo{year}{2005}).
\newblock \bibinfo{title}{{Chondrule-forming Shock Fronts in the Solar Nebula:
  A Possible Unified Scenario for Planet and Chondrite Formation}}.
\newblock {\it \bibinfo{journal}{\apjl}\/},  {\it \bibinfo{volume}{621}\/},
  \bibinfo{pages}{L137--L140}. \DOIprefix\doi{10.1086/429160}.
  \href{http://arxiv.org/abs/astro-ph/0501592}{\tt arXiv:astro-ph/0501592}.
\bibitem[{{Brearley}(1999)}]{1999Sci...285.1380B}
\bibinfo{author}{{Brearley}, A.~J.} (\bibinfo{year}{1999}).
\newblock \bibinfo{title}{{Origin of graphitic carbon and pentlandite in matrix
  olivines in the Allende meteorite.}}
\newblock {\it \bibinfo{journal}{Science}\/},  {\it \bibinfo{volume}{285}\/},
  \bibinfo{pages}{1380--1382}. \DOIprefix\doi{10.1126/science.285.5432.1380}.
\bibitem[{{Carballido} et~al.(2006){Carballido}, {Fromang} \&
  {Papaloizou}}]{2006MNRAS.373.1633C}
\bibinfo{author}{{Carballido}, A.}, \bibinfo{author}{{Fromang}, S.}, \&
  \bibinfo{author}{{Papaloizou}, J.} (\bibinfo{year}{2006}).
\newblock \bibinfo{title}{{Mid-plane sedimentation of large solid bodies in
  turbulent protoplanetary discs}}.
\newblock {\it \bibinfo{journal}{\mnras}\/},  {\it \bibinfo{volume}{373}\/},
  \bibinfo{pages}{1633--1640}.
  \DOIprefix\doi{10.1111/j.1365-2966.2006.11118.x}.
  \href{http://arxiv.org/abs/astro-ph/0610075}{\tt arXiv:astro-ph/0610075}.
\bibitem[{{Desch}(2007)}]{2007ApJ...671..878D}
\bibinfo{author}{{Desch}, S.~J.} (\bibinfo{year}{2007}).
\newblock \bibinfo{title}{{Mass Distribution and Planet Formation in the Solar
  Nebula}}.
\newblock {\it \bibinfo{journal}{\apj}\/},  {\it \bibinfo{volume}{671}\/},
  \bibinfo{pages}{878--893}. \DOIprefix\doi{10.1086/522825}.
\bibitem[{{Dubrulle} et~al.(1995){Dubrulle}, {Morfill} \&
  {Sterzik}}]{1995Icar..114..237D}
\bibinfo{author}{{Dubrulle}, B.}, \bibinfo{author}{{Morfill}, G.}, \&
  \bibinfo{author}{{Sterzik}, M.} (\bibinfo{year}{1995}).
\newblock \bibinfo{title}{{The dust subdisk in the protoplanetary nebula}}.
\newblock {\it \bibinfo{journal}{\icarus}\/},  {\it \bibinfo{volume}{114}\/},
  \bibinfo{pages}{237--246}. \DOIprefix\doi{10.1006/icar.1995.1058}.
\bibitem[{{Fedkin} \& {Grossman}(2013)}]{Fedkin13}
\bibinfo{author}{{Fedkin}, A.~V.}, \& \bibinfo{author}{{Grossman}, L.}
  (\bibinfo{year}{2013}).
\newblock \bibinfo{title}{{Vapor saturation of sodium: Key to unlocking the
  origin of chondrules}}.
\newblock {\it \bibinfo{journal}{\gca}\/},  {\it \bibinfo{volume}{112}\/},
  \bibinfo{pages}{226--250}. \DOIprefix\doi{10.1016/j.gca.2013.02.020}.
\bibitem[{{Flynn} et~al.(2013){Flynn}, {Wirick} \&
  {Keller}}]{2013EP&S...65.1159F}
\bibinfo{author}{{Flynn}, G.~J.}, \bibinfo{author}{{Wirick}, S.}, \&
  \bibinfo{author}{{Keller}, L.~P.} (\bibinfo{year}{2013}).
\newblock \bibinfo{title}{{Organic grain coatings in primitive interplanetary
  dust particles: Implications for grain sticking in the Solar Nebula}}.
\newblock {\it \bibinfo{journal}{Earth, Planets, and Space}\/},  {\it
  \bibinfo{volume}{65}\/}, \bibinfo{pages}{1159--1166}.
  \DOIprefix\doi{10.5047/eps.2013.05.007}.
\bibitem[{{Friedrich} et~al.(2014){Friedrich}, {Weisberg}, {Ebel}, {Biltz},
  {Corbett}, {Iotzov}, {Khan} \& {Wolman}}]{Friedrich2014}
\bibinfo{author}{{Friedrich}, J.~M.}, \bibinfo{author}{{Weisberg}, M.~K.},
  \bibinfo{author}{{Ebel}, D.~S.}, \bibinfo{author}{{Biltz}, A.~E.},
  \bibinfo{author}{{Corbett}, B.~M.}, \bibinfo{author}{{Iotzov}, I.~V.},
  \bibinfo{author}{{Khan}, W.~S.}, \& \bibinfo{author}{{Wolman}, M.~D.}
  (\bibinfo{year}{2014}).
\newblock \bibinfo{title}{{Chondrule size and related physical properties: a
  compilation and evaluation of current data across all meteorite groups}}.
\newblock {\it \bibinfo{journal}{ArXiv e-prints}\/}, .
  \href{http://arxiv.org/abs/1408.6581}{\tt arXiv:1408.6581}.
\bibitem[{{Fromang} \& {Papaloizou}(2006)}]{2006A&A...452..751F}
\bibinfo{author}{{Fromang}, S.}, \& \bibinfo{author}{{Papaloizou}, J.}
  (\bibinfo{year}{2006}).
\newblock \bibinfo{title}{{Dust settling in local simulations of turbulent
  protoplanetary disks}}.
\newblock {\it \bibinfo{journal}{\aap}\/},  {\it \bibinfo{volume}{452}\/},
  \bibinfo{pages}{751--762}. \DOIprefix\doi{10.1051/0004-6361:20054612}.
  \href{http://arxiv.org/abs/astro-ph/0603153}{\tt arXiv:astro-ph/0603153}.
\bibitem[{{Gammie}(1996)}]{1996ApJ...457..355G}
\bibinfo{author}{{Gammie}, C.~F.} (\bibinfo{year}{1996}).
\newblock \bibinfo{title}{{Layered Accretion in T Tauri Disks}}.
\newblock {\it \bibinfo{journal}{\apj}\/},  {\it \bibinfo{volume}{457}\/},
  \bibinfo{pages}{355}. \DOIprefix\doi{10.1086/176735}.
\bibitem[{{Grossman} \& {Brearley}(2005)}]{2005M&PS...40...87G}
\bibinfo{author}{{Grossman}, J.~N.}, \& \bibinfo{author}{{Brearley}, A.~J.}
  (\bibinfo{year}{2005}).
\newblock \bibinfo{title}{{The onset of metamorphism in ordinary and
  carbonaceous chondrites}}.
\newblock {\it \bibinfo{journal}{Meteoritics and Planetary Science}\/},  {\it
  \bibinfo{volume}{40}\/}, \bibinfo{pages}{87}.
  \DOIprefix\doi{10.1111/j.1945-5100.2005.tb00366.x}.
\bibitem[{{G{\"u}ttler} et~al.(2010){G{\"u}ttler}, {Blum}, {Zsom}, {Ormel} \&
  {Dullemond}}]{2010A&A...513A..56G}
\bibinfo{author}{{G{\"u}ttler}, C.}, \bibinfo{author}{{Blum}, J.},
  \bibinfo{author}{{Zsom}, A.}, \bibinfo{author}{{Ormel}, C.~W.}, \&
  \bibinfo{author}{{Dullemond}, C.~P.} (\bibinfo{year}{2010}).
\newblock \bibinfo{title}{{The outcome of protoplanetary dust growth: pebbles,
  boulders, or planetesimals?. I. Mapping the zoo of laboratory collision
  experiments}}.
\newblock {\it \bibinfo{journal}{\aap}\/},  {\it \bibinfo{volume}{513}\/},
  \bibinfo{pages}{A56}. \DOIprefix\doi{10.1051/0004-6361/200912852}.
  \href{http://arxiv.org/abs/0910.4251}{\tt arXiv:0910.4251}.
\bibitem[{{Hartmann} \& {Kenyon}(1996)}]{1996ARA&A..34..207H}
\bibinfo{author}{{Hartmann}, L.}, \& \bibinfo{author}{{Kenyon}, S.~J.}
  (\bibinfo{year}{1996}).
\newblock \bibinfo{title}{{The FU Orionis Phenomenon}}.
\newblock {\it \bibinfo{journal}{\araa}\/},  {\it \bibinfo{volume}{34}\/},
  \bibinfo{pages}{207--240}. \DOIprefix\doi{10.1146/annurev.astro.34.1.207}.
\bibitem[{{Hayashi}(1981)}]{1981PThPS..70...35H}
\bibinfo{author}{{Hayashi}, C.} (\bibinfo{year}{1981}).
\newblock \bibinfo{title}{{Structure of the Solar Nebula, Growth and Decay of
  Magnetic Fields and Effects of Magnetic and Turbulent Viscosities on the
  Nebula}}.
\newblock {\it \bibinfo{journal}{Progress of Theoretical Physics
  Supplement}\/},  {\it \bibinfo{volume}{70}\/}, \bibinfo{pages}{35--53}.
  \DOIprefix\doi{10.1143/PTPS.70.35}.
\bibitem[{{Hewins} \& {Radomsky}(1990)}]{1990Metic..25..309H}
\bibinfo{author}{{Hewins}, R.~H.}, \& \bibinfo{author}{{Radomsky}, P.~M.}
  (\bibinfo{year}{1990}).
\newblock \bibinfo{title}{{Temperature conditions for chondrule formation}}.
\newblock {\it \bibinfo{journal}{Meteoritics}\/},  {\it
  \bibinfo{volume}{25}\/}, \bibinfo{pages}{309--318}.
\bibitem[{{Hewins} et~al.(2012){Hewins}, {Zanda} \& {Bendersky}}]{Hewins12}
\bibinfo{author}{{Hewins}, R.~H.}, \bibinfo{author}{{Zanda}, B.}, \&
  \bibinfo{author}{{Bendersky}, C.} (\bibinfo{year}{2012}).
\newblock \bibinfo{title}{{Evaporation and recondensation of sodium in
  Semarkona Type II chondrules}}.
\newblock {\it \bibinfo{journal}{\gca}\/},  {\it \bibinfo{volume}{78}\/},
  \bibinfo{pages}{1--17}. \DOIprefix\doi{10.1016/j.gca.2011.11.027}.
\bibitem[{{Hezel} \& {Palme}(2010)}]{2010E&PSL.294...85H}
\bibinfo{author}{{Hezel}, D.~C.}, \& \bibinfo{author}{{Palme}, H.}
  (\bibinfo{year}{2010}).
\newblock \bibinfo{title}{{The chemical relationship between chondrules and
  matrix and the chondrule matrix complementarity}}.
\newblock {\it \bibinfo{journal}{Earth and Planetary Science Letters}\/},  {\it
  \bibinfo{volume}{294}\/}, \bibinfo{pages}{85--93}.
  \DOIprefix\doi{10.1016/j.epsl.2010.03.008}.
\bibitem[{{Hood}(1998)}]{1998M&PS...33...97H}
\bibinfo{author}{{Hood}, L.~L.} (\bibinfo{year}{1998}).
\newblock \bibinfo{title}{{Thermal processing of chondrule and CAI precursors
  in planetesimal bow shocks}}.
\newblock {\it \bibinfo{journal}{Meteoritics and Planetary Science}\/},  {\it
  \bibinfo{volume}{33}\/}, \bibinfo{pages}{97--107}.
  \DOIprefix\doi{10.1111/j.1945-5100.1998.tb01611.x}.
\bibitem[{{Hubbard}(2012)}]{2012MNRAS.426..784H}
\bibinfo{author}{{Hubbard}, A.} (\bibinfo{year}{2012}).
\newblock \bibinfo{title}{{Turbulence-induced collisional velocities and
  density enhancements: large inertial range results from shell models}}.
\newblock {\it \bibinfo{journal}{\mnras}\/},  {\it \bibinfo{volume}{426}\/},
  \bibinfo{pages}{784--795}. \DOIprefix\doi{10.1111/j.1365-2966.2012.21758.x}.
  \href{http://arxiv.org/abs/1207.5365}{\tt arXiv:1207.5365}.
\bibitem[{{Hubbard}(2013)}]{2013MNRAS.432.1274H}
\bibinfo{author}{{Hubbard}, A.} (\bibinfo{year}{2013}).
\newblock \bibinfo{title}{{Turbulence-induced collision velocities and rates
  between different sized dust grains}}.
\newblock {\it \bibinfo{journal}{\mnras}\/},  {\it \bibinfo{volume}{432}\/},
  \bibinfo{pages}{1274--1284}. \DOIprefix\doi{10.1093/mnras/stt543}.
  \href{http://arxiv.org/abs/1303.6639}{\tt arXiv:1303.6639}.
\bibitem[{{Hubbard} \& {Ebel}(2014)}]{2014Icar..237...84H}
\bibinfo{author}{{Hubbard}, A.}, \& \bibinfo{author}{{Ebel}, D.~S.}
  (\bibinfo{year}{2014}).
\newblock \bibinfo{title}{{Protoplanetary dust porosity and FU Orionis
  outbursts: Solving the mystery of Earth's missing volatiles}}.
\newblock {\it \bibinfo{journal}{\icarus}\/},  {\it \bibinfo{volume}{237}\/},
  \bibinfo{pages}{84--96}. \DOIprefix\doi{10.1016/j.icarus.2014.04.015}.
  \href{http://arxiv.org/abs/1404.3995}{\tt arXiv:1404.3995}.
\bibitem[{{Hubbard} et~al.(2012){Hubbard}, {McNally} \& {Mac
  Low}}]{2012ApJ...761...58H}
\bibinfo{author}{{Hubbard}, A.}, \bibinfo{author}{{McNally}, C.~P.}, \&
  \bibinfo{author}{{Mac Low}, M.-M.} (\bibinfo{year}{2012}).
\newblock \bibinfo{title}{{Short Circuits in Thermally Ionized Plasmas: A
  Mechanism for Intermittent Heating of Protoplanetary Disks}}.
\newblock {\it \bibinfo{journal}{\apj}\/},  {\it \bibinfo{volume}{761}\/},
  \bibinfo{pages}{58}. \DOIprefix\doi{10.1088/0004-637X/761/1/58}.
  \href{http://arxiv.org/abs/1206.7096}{\tt arXiv:1206.7096}.
\bibitem[{{Huss} \& {Lewis}(1994)}]{1994Metic..29..811H}
\bibinfo{author}{{Huss}, G.~R.}, \& \bibinfo{author}{{Lewis}, R.~S.}
  (\bibinfo{year}{1994}).
\newblock \bibinfo{title}{{Noble gases in presolar diamonds II: Component
  abundances reflect thermal processing}}.
\newblock {\it \bibinfo{journal}{Meteoritics}\/},  {\it
  \bibinfo{volume}{29}\/}, \bibinfo{pages}{811--829}.
\bibitem[{{Ingleby} et~al.(2013){Ingleby}, {Calvet}, {Herczeg}, {Blaty},
  {Walter}, {Ardila}, {Alexander}, {Edwards}, {Espaillat}, {Gregory},
  {Hillenbrand} \& {Brown}}]{2013ApJ...767..112I}
\bibinfo{author}{{Ingleby}, L.}, \bibinfo{author}{{Calvet}, N.},
  \bibinfo{author}{{Herczeg}, G.}, \bibinfo{author}{{Blaty}, A.},
  \bibinfo{author}{{Walter}, F.}, \bibinfo{author}{{Ardila}, D.},
  \bibinfo{author}{{Alexander}, R.}, \bibinfo{author}{{Edwards}, S.},
  \bibinfo{author}{{Espaillat}, C.}, \bibinfo{author}{{Gregory}, S.~G.},
  \bibinfo{author}{{Hillenbrand}, L.}, \& \bibinfo{author}{{Brown}, A.}
  (\bibinfo{year}{2013}).
\newblock \bibinfo{title}{{Accretion Rates for T Tauri Stars Using Nearly
  Simultaneous Ultraviolet and Optical Spectra}}.
\newblock {\it \bibinfo{journal}{\apj}\/},  {\it \bibinfo{volume}{767}\/},
  \bibinfo{pages}{112}. \DOIprefix\doi{10.1088/0004-637X/767/2/112}.
  \href{http://arxiv.org/abs/1303.0769}{\tt arXiv:1303.0769}.
\bibitem[{{Jacquet}(2014)}]{2014Icar..232..176J}
\bibinfo{author}{{Jacquet}, E.} (\bibinfo{year}{2014}).
\newblock \bibinfo{title}{{The quasi-universality of chondrule size as a
  constraint for chondrule formation models}}.
\newblock {\it \bibinfo{journal}{\icarus}\/},  {\it \bibinfo{volume}{232}\/},
  \bibinfo{pages}{176--186}. \DOIprefix\doi{10.1016/j.icarus.2014.01.012}.
  \href{http://arxiv.org/abs/1401.3721}{\tt arXiv:1401.3721}.
\bibitem[{{Johansen} et~al.(2007){Johansen}, {Oishi}, {Mac Low}, {Klahr},
  {Henning} \& {Youdin}}]{2007Natur.448.1022J}
\bibinfo{author}{{Johansen}, A.}, \bibinfo{author}{{Oishi}, J.~S.},
  \bibinfo{author}{{Mac Low}, M.-M.}, \bibinfo{author}{{Klahr}, H.},
  \bibinfo{author}{{Henning}, T.}, \& \bibinfo{author}{{Youdin}, A.}
  (\bibinfo{year}{2007}).
\newblock \bibinfo{title}{{Rapid planetesimal formation in turbulent
  circumstellar disks}}.
\newblock {\it \bibinfo{journal}{\nat}\/},  {\it \bibinfo{volume}{448}\/},
  \bibinfo{pages}{1022--1025}. \DOIprefix\doi{10.1038/nature06086}.
  \href{http://arxiv.org/abs/0708.3890}{\tt arXiv:0708.3890}.
\bibitem[{{Johansen} et~al.(2009){Johansen}, {Youdin} \& {Mac
  Low}}]{2009ApJ...704L..75J}
\bibinfo{author}{{Johansen}, A.}, \bibinfo{author}{{Youdin}, A.}, \&
  \bibinfo{author}{{Mac Low}, M.-M.} (\bibinfo{year}{2009}).
\newblock \bibinfo{title}{{Particle Clumping and Planetesimal Formation Depend
  Strongly on Metallicity}}.
\newblock {\it \bibinfo{journal}{\apjl}\/},  {\it \bibinfo{volume}{704}\/},
  \bibinfo{pages}{L75--L79}. \DOIprefix\doi{10.1088/0004-637X/704/2/L75}.
  \href{http://arxiv.org/abs/0909.0259}{\tt arXiv:0909.0259}.
\bibitem[{{Jones}(2012)}]{2012M&PS...47.1176J}
\bibinfo{author}{{Jones}, R.~H.} (\bibinfo{year}{2012}).
\newblock \bibinfo{title}{{Petrographic constraints on the diversity of
  chondrule reservoirs in the protoplanetary disk}}.
\newblock {\it \bibinfo{journal}{Meteoritics and Planetary Science}\/},  {\it
  \bibinfo{volume}{47}\/}, \bibinfo{pages}{1176--1190}.
  \DOIprefix\doi{10.1111/j.1945-5100.2011.01327.x}.
\bibitem[{{Lee} et~al.(2010){Lee}, {Chiang}, {Asay-Davis} \&
  {Barranco}}]{2010ApJ...725.1938L}
\bibinfo{author}{{Lee}, A.~T.}, \bibinfo{author}{{Chiang}, E.},
  \bibinfo{author}{{Asay-Davis}, X.}, \& \bibinfo{author}{{Barranco}, J.}
  (\bibinfo{year}{2010}).
\newblock \bibinfo{title}{{Forming Planetesimals by Gravitational Instability.
  II. How Dust Settles to its Marginally Stable State}}.
\newblock {\it \bibinfo{journal}{\apj}\/},  {\it \bibinfo{volume}{725}\/},
  \bibinfo{pages}{1938--1954}. \DOIprefix\doi{10.1088/0004-637X/725/2/1938}.
  \href{http://arxiv.org/abs/1010.0250}{\tt arXiv:1010.0250}.
\bibitem[{{Lobo} et~al.(2014){Lobo}, {Wallace} \& {Ebel}}]{2014LPI....45.1423L}
\bibinfo{author}{{Lobo}, A.}, \bibinfo{author}{{Wallace}, S.~W.}, \&
  \bibinfo{author}{{Ebel}, D.~S.} (\bibinfo{year}{2014}).
\newblock \bibinfo{title}{{Modal Abundances, Chemistry and Sizes of Clasts in
  the Semarkona (LL3.0) Chondrite by X-Ray Map Analysis}}.
\newblock In {\it \bibinfo{booktitle}{Lunar and Planetary Science
  Conference}\/} (p. \bibinfo{pages}{1423}).
\newblock volume~\bibinfo{volume}{45} of {\it \bibinfo{series}{Lunar and
  Planetary Science Conference}\/}.
\bibitem[{{Lodders}(2003)}]{2003ApJ...591.1220L}
\bibinfo{author}{{Lodders}, K.} (\bibinfo{year}{2003}).
\newblock \bibinfo{title}{{Solar System Abundances and Condensation
  Temperatures of the Elements}}.
\newblock {\it \bibinfo{journal}{\apj}\/},  {\it \bibinfo{volume}{591}\/},
  \bibinfo{pages}{1220--1247}. \DOIprefix\doi{10.1086/375492}.
\bibitem[{{McNally} et~al.(2013){McNally}, {Hubbard}, {Mac Low}, {Ebel} \&
  {D'Alessio}}]{2013ApJ...767L...2M}
\bibinfo{author}{{McNally}, C.~P.}, \bibinfo{author}{{Hubbard}, A.},
  \bibinfo{author}{{Mac Low}, M.-M.}, \bibinfo{author}{{Ebel}, D.~S.}, \&
  \bibinfo{author}{{D'Alessio}, P.} (\bibinfo{year}{2013}).
\newblock \bibinfo{title}{{Mineral Processing by Short Circuits in
  Protoplanetary Disks}}.
\newblock {\it \bibinfo{journal}{\apjl}\/},  {\it \bibinfo{volume}{767}\/},
  \bibinfo{pages}{L2}. \DOIprefix\doi{10.1088/2041-8205/767/1/L2}.
  \href{http://arxiv.org/abs/1301.1698}{\tt arXiv:1301.1698}.
\bibitem[{{McNally} et~al.(2014){McNally}, {Hubbard}, {Yang} \& {Mac
  Low}}]{2014ApJ...791...62M}
\bibinfo{author}{{McNally}, C.~P.}, \bibinfo{author}{{Hubbard}, A.},
  \bibinfo{author}{{Yang}, C.-C.}, \& \bibinfo{author}{{Mac Low}, M.-M.}
  (\bibinfo{year}{2014}).
\newblock \bibinfo{title}{{Temperature Fluctuations Driven by Magnetorotational
  Instability in Protoplanetary Disks}}.
\newblock {\it \bibinfo{journal}{\apj}\/},  {\it \bibinfo{volume}{791}\/},
  \bibinfo{pages}{62}. \DOIprefix\doi{10.1088/0004-637X/791/1/62}.
  \href{http://arxiv.org/abs/1406.5195}{\tt arXiv:1406.5195}.
\bibitem[{{Metzler}(2012)}]{2012M&PS...47.2193M}
\bibinfo{author}{{Metzler}, K.} (\bibinfo{year}{2012}).
\newblock \bibinfo{title}{{Ultrarapid chondrite formation by hot chondrule
  accretion? Evidence from unequilibrated ordinary chondrites}}.
\newblock {\it \bibinfo{journal}{Meteoritics and Planetary Science}\/},  {\it
  \bibinfo{volume}{47}\/}, \bibinfo{pages}{2193--2217}.
  \DOIprefix\doi{10.1111/j.1945-5100.2012.01412.x}.
\bibitem[{{Morris} et~al.(2012){Morris}, {Boley}, {Desch} \&
  {Athanassiadou}}]{2012ApJ...752...27M}
\bibinfo{author}{{Morris}, M.~A.}, \bibinfo{author}{{Boley}, A.~C.},
  \bibinfo{author}{{Desch}, S.~J.}, \& \bibinfo{author}{{Athanassiadou}, T.}
  (\bibinfo{year}{2012}).
\newblock \bibinfo{title}{{Chondrule Formation in Bow Shocks around Eccentric
  Planetary Embryos}}.
\newblock {\it \bibinfo{journal}{\apj}\/},  {\it \bibinfo{volume}{752}\/},
  \bibinfo{pages}{27}. \DOIprefix\doi{10.1088/0004-637X/752/1/27}.
  \href{http://arxiv.org/abs/1204.0739}{\tt arXiv:1204.0739}.
\bibitem[{{Ormel} et~al.(2007){Ormel}, {Spaans} \&
  {Tielens}}]{2007A&A...461..215O}
\bibinfo{author}{{Ormel}, C.~W.}, \bibinfo{author}{{Spaans}, M.}, \&
  \bibinfo{author}{{Tielens}, A.~G.~G.~M.} (\bibinfo{year}{2007}).
\newblock \bibinfo{title}{{Dust coagulation in protoplanetary disks: porosity
  matters}}.
\newblock {\it \bibinfo{journal}{\aap}\/},  {\it \bibinfo{volume}{461}\/},
  \bibinfo{pages}{215--232}. \DOIprefix\doi{10.1051/0004-6361:20065949}.
  \href{http://arxiv.org/abs/astro-ph/0610030}{\tt arXiv:astro-ph/0610030}.
\bibitem[{{Pan} \& {Padoan}(2013)}]{2013ApJ...776...12P}
\bibinfo{author}{{Pan}, L.}, \& \bibinfo{author}{{Padoan}, P.}
  (\bibinfo{year}{2013}).
\newblock \bibinfo{title}{{Turbulence-induced Relative Velocity of Dust
  Particles. I. Identical Particles}}.
\newblock {\it \bibinfo{journal}{\apj}\/},  {\it \bibinfo{volume}{776}\/},
  \bibinfo{pages}{12}. \DOIprefix\doi{10.1088/0004-637X/776/1/12}.
  \href{http://arxiv.org/abs/1305.0307}{\tt arXiv:1305.0307}.
\bibitem[{{Pan} et~al.(2014){Pan}, {Padoan} \& {Scalo}}]{2014ApJ...791...48P}
\bibinfo{author}{{Pan}, L.}, \bibinfo{author}{{Padoan}, P.}, \&
  \bibinfo{author}{{Scalo}, J.} (\bibinfo{year}{2014}).
\newblock \bibinfo{title}{{Turbulence-induced Relative Velocity of Dust
  Particles. II. The Bidisperse Case}}.
\newblock {\it \bibinfo{journal}{\apj}\/},  {\it \bibinfo{volume}{791}\/},
  \bibinfo{pages}{48}. \DOIprefix\doi{10.1088/0004-637X/791/1/48}.
  \href{http://arxiv.org/abs/1403.3865}{\tt arXiv:1403.3865}.
\bibitem[{{Rubin}(2013)}]{2013M&PS...48..445R}
\bibinfo{author}{{Rubin}, A.~E.} (\bibinfo{year}{2013}).
\newblock \bibinfo{title}{{Multiple melting in a four-layered barred-olivine
  chondrule with compositionally heterogeneous glass from LL3.0 Semarkona}}.
\newblock {\it \bibinfo{journal}{Meteoritics and Planetary Science}\/},  {\it
  \bibinfo{volume}{48}\/}, \bibinfo{pages}{445--456}.
  \DOIprefix\doi{10.1111/maps.12069}.
\bibitem[{{Shakura} \& {Sunyaev}(1973)}]{1973A&A....24..337S}
\bibinfo{author}{{Shakura}, N.~I.}, \& \bibinfo{author}{{Sunyaev}, R.~A.}
  (\bibinfo{year}{1973}).
\newblock \bibinfo{title}{{Black holes in binary systems. Observational
  appearance.}}
\newblock {\it \bibinfo{journal}{\aap}\/},  {\it \bibinfo{volume}{24}\/},
  \bibinfo{pages}{337--355}.
\bibitem[{{Susa} \& {Nakamoto}(2002)}]{2002ApJ...564L..57S}
\bibinfo{author}{{Susa}, H.}, \& \bibinfo{author}{{Nakamoto}, T.}
  (\bibinfo{year}{2002}).
\newblock \bibinfo{title}{{On the Maximal Size of Chondrules in Shock Wave
  Heating Model}}.
\newblock {\it \bibinfo{journal}{\apjl}\/},  {\it \bibinfo{volume}{564}\/},
  \bibinfo{pages}{L57--L60}. \DOIprefix\doi{10.1086/338789}.
\bibitem[{{Takeuchi} \& {Lin}(2002)}]{2002ApJ...581.1344T}
\bibinfo{author}{{Takeuchi}, T.}, \& \bibinfo{author}{{Lin}, D.~N.~C.}
  (\bibinfo{year}{2002}).
\newblock \bibinfo{title}{{Radial Flow of Dust Particles in Accretion Disks}}.
\newblock {\it \bibinfo{journal}{\apj}\/},  {\it \bibinfo{volume}{581}\/},
  \bibinfo{pages}{1344--1355}. \DOIprefix\doi{10.1086/344437}.
  \href{http://arxiv.org/abs/astro-ph/0208552}{\tt arXiv:astro-ph/0208552}.
\bibitem[{{Voelk} et~al.(1980){Voelk}, {Jones}, {Morfill} \&
  {Roeser}}]{1980A&A....85..316V}
\bibinfo{author}{{Voelk}, H.~J.}, \bibinfo{author}{{Jones}, F.~C.},
  \bibinfo{author}{{Morfill}, G.~E.}, \& \bibinfo{author}{{Roeser}, S.}
  (\bibinfo{year}{1980}).
\newblock \bibinfo{title}{{Collisions between grains in a turbulent gas}}.
\newblock {\it \bibinfo{journal}{\aap}\/},  {\it \bibinfo{volume}{85}\/},
  \bibinfo{pages}{316--325}.
\bibitem[{{Weidenschilling}(1977)}]{1977MNRAS.180...57W}
\bibinfo{author}{{Weidenschilling}, S.~J.} (\bibinfo{year}{1977}).
\newblock \bibinfo{title}{{Aerodynamics of solid bodies in the solar nebula}}.
\newblock {\it \bibinfo{journal}{\mnras}\/},  {\it \bibinfo{volume}{180}\/},
  \bibinfo{pages}{57--70}.
\bibitem[{{Weidenschilling}(1980)}]{1980Icar...44..172W}
\bibinfo{author}{{Weidenschilling}, S.~J.} (\bibinfo{year}{1980}).
\newblock \bibinfo{title}{{Dust to planetesimals - Settling and coagulation in
  the solar nebula}}.
\newblock {\it \bibinfo{journal}{\icarus}\/},  {\it \bibinfo{volume}{44}\/},
  \bibinfo{pages}{172--189}. \DOIprefix\doi{10.1016/0019-1035(80)90064-0}.
\bibitem[{{Weidenschilling}(2006)}]{2006Icar..181..572W}
\bibinfo{author}{{Weidenschilling}, S.~J.} (\bibinfo{year}{2006}).
\newblock \bibinfo{title}{{Models of particle layers in the midplane of the
  solar nebula}}.
\newblock {\it \bibinfo{journal}{\icarus}\/},  {\it \bibinfo{volume}{181}\/},
  \bibinfo{pages}{572--586}. \DOIprefix\doi{10.1016/j.icarus.2005.11.017}.
\bibitem[{{Weisberg} et~al.(2006){Weisberg}, {McCoy} \&
  {Krot}}]{2006mess.book...19W}
\bibinfo{author}{{Weisberg}, M.~K.}, \bibinfo{author}{{McCoy}, T.~J.}, \&
  \bibinfo{author}{{Krot}, A.~N.} (\bibinfo{year}{2006}).
\newblock \bibinfo{title}{{Systematics and Evaluation of Meteorite
  Classification}}.
\newblock In \bibinfo{editor}{D.~S. {Lauretta}}, \& \bibinfo{editor}{H.~Y.
  {McSween}} (Eds.), {\it \bibinfo{booktitle}{Meteorites and the Early Solar
  System II}\/} (pp. \bibinfo{pages}{19--52}).
\bibitem[{{Weisberg} \& {Prinz}(1996)}]{1996cpd..conf..119W}
\bibinfo{author}{{Weisberg}, M.~K.}, \& \bibinfo{author}{{Prinz}, M.}
  (\bibinfo{year}{1996}).
\newblock \bibinfo{title}{{Agglomeratic chondrules, chondrule precursors, and
  incomplete melting.}}
\newblock In \bibinfo{editor}{R.~H. {Hewins}}, \bibinfo{editor}{R.~H. {Jones}},
  \& \bibinfo{editor}{E.~R.~D. {Scott}} (Eds.), {\it
  \bibinfo{booktitle}{Chondrules and the Protoplanetary Disk}\/} (pp.
  \bibinfo{pages}{119--127}).
\bibitem[{{Windmark} et~al.(2012){Windmark}, {Birnstiel}, {Ormel} \&
  {Dullemond}}]{2012A&A...544L..16W}
\bibinfo{author}{{Windmark}, F.}, \bibinfo{author}{{Birnstiel}, T.},
  \bibinfo{author}{{Ormel}, C.~W.}, \& \bibinfo{author}{{Dullemond}, C.~P.}
  (\bibinfo{year}{2012}).
\newblock \bibinfo{title}{{Breaking through: The effects of a velocity
  distribution on barriers to dust growth}}.
\newblock {\it \bibinfo{journal}{\aap}\/},  {\it \bibinfo{volume}{544}\/},
  \bibinfo{pages}{L16}. \DOIprefix\doi{10.1051/0004-6361/201220004}.
  \href{http://arxiv.org/abs/1208.0304}{\tt arXiv:1208.0304}.
\bibitem[{{Youdin} \& {Goodman}(2005)}]{2005ApJ...620..459Y}
\bibinfo{author}{{Youdin}, A.~N.}, \& \bibinfo{author}{{Goodman}, J.}
  (\bibinfo{year}{2005}).
\newblock \bibinfo{title}{{Streaming Instabilities in Protoplanetary Disks}}.
\newblock {\it \bibinfo{journal}{\apj}\/},  {\it \bibinfo{volume}{620}\/},
  \bibinfo{pages}{459--469}. \DOIprefix\doi{10.1086/426895}.
  \href{http://arxiv.org/abs/astro-ph/0409263}{\tt arXiv:astro-ph/0409263}.
\bibitem[{{Zsom} et~al.(2010){Zsom}, {Ormel}, {G{\"u}ttler}, {Blum} \&
  {Dullemond}}]{2010A&A...513A..57Z}
\bibinfo{author}{{Zsom}, A.}, \bibinfo{author}{{Ormel}, C.~W.},
  \bibinfo{author}{{G{\"u}ttler}, C.}, \bibinfo{author}{{Blum}, J.}, \&
  \bibinfo{author}{{Dullemond}, C.~P.} (\bibinfo{year}{2010}).
\newblock \bibinfo{title}{{The outcome of protoplanetary dust growth: pebbles,
  boulders, or planetesimals? II. Introducing the bouncing barrier}}.
\newblock {\it \bibinfo{journal}{\aap}\/},  {\it \bibinfo{volume}{513}\/},
  \bibinfo{pages}{A57}. \DOIprefix\doi{10.1051/0004-6361/200912976}.
  \href{http://arxiv.org/abs/1001.0488}{\tt arXiv:1001.0488}.

\end{thebibliography}

\end{document}